\documentclass[11pt,a4paper]{article}

\usepackage[T1]{fontenc}
\usepackage[utf8]{inputenc}
\usepackage{lmodern}
\usepackage{microtype}
\usepackage[a4paper,margin=28mm]{geometry}
\usepackage{amsmath,amssymb,amsthm,mathtools,todonotes}
\usepackage{booktabs}
\usepackage{graphicx,subcaption}
\usepackage{aliascnt}
\usepackage[all,defaultlines=4]{nowidow}
\usepackage{tikz}
\usetikzlibrary{positioning,fit,calc,shapes.geometric}
\usepackage[numbers,sort&compress]{natbib}
\usepackage{hyperref}
\usepackage[nameinlink,capitalise,noabbrev]{cleveref}

\definecolor{linkblue}{RGB}{20,75,130}
\definecolor{darkred}{RGB}{150,30,35}
\definecolor{landmarkblue}{RGB}{45,105,165}
\hypersetup{
  colorlinks=true,
  linkcolor=linkblue,
  citecolor=linkblue,
  urlcolor=linkblue,
  pdftitle={Metric Dimension Parameterized by Bag-Connected Treewidth and Bag-Connected Pathwidth},
  pdfauthor={Dibyayan Chakraborty, Sebastian Ordyniak, and Sebastian Siebertz}
}

\newtheorem{theorem}{Theorem}[section]
\newaliascnt{lemma}{theorem}
\newtheorem{lemma}[lemma]{Lemma}
\aliascntresetthe{lemma}
\newaliascnt{corollary}{theorem}

\aliascntresetthe{corollary}

\newcommand{\dist}{\operatorname{dist}}
\newcommand{\tw}{\operatorname{tw}}
\newcommand{\ctw}{\operatorname{ctw}}
\newcommand{\cpw}{\operatorname{gs-cpw}}
\newcommand{\bcpw}{\operatorname{cpw}}
\newcommand{\tl}{\operatorname{tl}}
\newcommand{\mdim}{\operatorname{dim}}
\newcommand{\gammaiso}{\gamma_{\mathrm{iso}}}
\newcommand{\MetricDimension}{\textnormal{\textsc{Metric Dimension}}}
\newcommand{\VertexCover}{\textnormal{\textsc{Vertex Cover}}}

\title{Metric Dimension Parameterized by Bag-Connected Treewidth and Bag-Connected Pathwidth\thanks{\textbf{Declaration of AI use.} While working on our paper on metric extensions of monadic second-order logic~\cite{ChakrabortyOrdyniakSiebertz2026MetricExtensions}, we considered the metric dimension problem studied here. After a few short discussions with us, the AI model Astra resolved the problem. We here present the results in a human-readable form and take full responsibility for the content of the paper.}}
\author{Dibyayan Chakraborty, Sebastian Ordyniak, and Sebastian Siebertz}
\date{October 2026}

\begin{document}
\maketitle

\begin{abstract}
The metric dimension of a graph is the minimum size of a set of vertices whose distance vectors distinguish all vertices of the graph.
We prove that \MetricDimension{} is NP-complete on planar graphs of treewidth and bag-connected treewidth exactly three.
The graphs in our construction have no induced cycle longer than four and have (prefix-connected) pathwidth at most four.
In particular, the problem admits no XP algorithm parameterized by bag-connected treewidth unless $\mathrm P=\mathrm{NP}$.
This answers questions of Bonnet and Purohit and of Bousquet, Deschamps, and Parreau.
We prove that the problem is fixed-parameter tractable when parameterized by bag-connected pathwidth. 
\end{abstract}

\section{Introduction}

Let $G$ be a finite, simple, connected graph.
A set $S\subseteq V(G)$ is a \emph{resolving set} if, for every two distinct vertices $u,v$, there is a vertex $s\in S$ with $\dist_G(s,u)\neq\dist_G(s,v)$.
The \emph{metric dimension} $\mdim(G)$ is the minimum size of a resolving set.
This notion was introduced independently by Slater and by Harary and Melter~\cite{Slater1975,HararyMelter1976}.
In the decision problem \MetricDimension{}, we are given a graph $G$ and an integer $k$ and ask whether $\mdim(G)\leq k$.

The problem is NP-complete in general~\cite[Problem~GT61]{GareyJohnson1979} and W[2]-complete when parameterized by solution size, even on subcubic graphs~\cite{HartungNichterlein2013}.
It is polynomial-time solvable on trees and outerplanar graphs~\cite{KhullerRaghavachariRosenfeld1996,DiazPottonenSernaVanLeeuwen2017}.
Belmonte, Fomin, Golovach, and Ramanujan proved fixed-parameter tractability for the combined parameter tree-length plus maximum degree~\cite{BelmonteFominGolovachRamanujan2017}.
Their result also gives fixed-parameter tractability for tree-length plus solution size.

For ordinary treewidth, Bonnet and Purohit proved W[1]-hardness, even for the larger parameter pathwidth and on graphs of constant maximum degree~\cite{BonnetPurohit2021}.
Li and Pilipczuk subsequently proved NP-hardness on graphs of treewidth and pathwidth at most $24$~\cite{LiPilipczuk2022}. See \cite{FoucaudEtAl2025VertexCover,FoucaudEtAl2024DoubleExponential,GalbyEtAl2023FeedbackVertexSet,FoucaudEtAl2017IntervalPermutation,DaillyFoucaudHakanen2026Digraphs,Tale2026StrongMetricDimension} for more results on this topic.
A recent preprint of Benoit gives NP-hardness on graphs of treewidth two and pathwidth at most five~\cite{Benoit2026}.

A tree decomposition is \emph{bag-connected} if every bag induces a connected subgraph.
The \emph{bag-connected treewidth} $\ctw(G)$ of a graph $G$ is the minimum width of such a decomposition~\cite{DiestelMuller2018,HamannWeissauer2016,jegou2014bag,JegouTerrioux2017Combining}.
A connected bag of size at most $q+1$ has diameter at most $q$, so bag-connected treewidth bounds both the size and the diameter of the bags.
Bonnet and Purohit asked whether \MetricDimension{} is fixed-parameter tractable when parameterized by bag-connected treewidth alone~\cite{BonnetPurohit2021}.
The algorithm for tree-length plus maximum degree already gives fixed-parameter tractability for bag-connected treewidth plus maximum degree, so the remaining question was whether the degree bound can be dropped.
Bousquet, Deschamps, and Parreau~\cite{BousquetDeschampsParreau2023} proved that \MetricDimension{} is fixed-parameter tractable by treewidth on chordal graphs.
They asked whether this extends to decompositions in which both the bag size and the bag diameter are bounded.
A positive answer would in particular imply fixed-parameter tractability by bag-connected treewidth.
We answer both questions negatively, even for planar graphs and even if one asks only for an XP algorithm, by showing that the problem is para-NP hard when parameterized by bag-connected treewidth. 

\begin{theorem}\label{thm:main}
\MetricDimension{} is NP-complete on graphs $G$ satisfying all of the following properties:
\vspace{-1mm}
\begin{itemize}
\setlength{\itemsep}{1pt}
  \item $G$ is planar.
  \item $\tw(G)=\ctw(G)=3$.
  \item $G$ has no induced cycle of length greater than four.
  \item $G$ admits a bag-connected tree decomposition of width three in which every bag has induced diameter at most two.
  \item $G$ admits a path decomposition of width at most four in which the union of every initial sequence of bags is connected.
\end{itemize}
The hardness holds even when both decompositions are given as part of the input.
\end{theorem}

Recall that an XP algorithm parameterized by bag-connected treewidth $q$ has running time $f(q)n^{g(q)}$ for computable functions $f,g$.
Such an algorithm runs in polynomial time when~$q=3$.
Hence \cref{thm:main} rules out an XP algorithm unless $\mathrm P=\mathrm{NP}$.

Our construction shows that the chordal result does not extend to $4$-chordal graphs, that is, graphs with no induced cycle longer than four.
The output graphs have tree-length two, whereas chordal graphs have tree-length at most one.
Furthermore, every isometric cycle in an output graph has length at most four.
Consequently, \MetricDimension{} is NP-hard even when both the treewidth and the maximum length of an isometric cycle are fixed constants.
\smallskip

% The idea of the proof is as follows.
% For every vertex $a$ of a \VertexCover{} instance~$F$, we construct a ladder, which we call a tail, attached to a common edge $pq$.
% Every edge $ab$ of $F$ receives a distinct integer level $h$.
% We place a marker at level $h$ in each of the tails corresponding to $a$ and $b$.
% Both markers have distance $h$ from $p$ and from $q$, so no vertex outside these two tails distinguishes them.
% A landmark at the beginning of either ladder does distinguish them.

% We must also resolve all the other pairs of vertices without distinguishing the marker pairs, and the number of landmarks needed for this should be independent of the vertex cover.
% For this purpose, we add a small graph at the beginning of each ladder and force two local landmarks by pairs of leaves with a common neighbor.
% The local landmarks distinguish corresponding rail vertices in different tails, but have the same distance to every marker at a given level.
% Once these landmarks have been chosen, choosing one further landmark in a tail corresponds to choosing the associated vertex for a vertex cover.
% We will prove that $\mdim(H_F)=2|V(F)|+2+\tau(F)$.
% \medskip

We next ask what happens if the decomposition tree is required to be a path.
The \emph{bag-connected pathwidth} $\bcpw(G)$ of a graph $G$ is the minimum width of a path decomposition in which every bag induces a connected subgraph.
There is a version of connected pathwidth studied in graph searching, where the unions of initial sequences of bags must be  connected~\cite{Dereniowski2012}. We denote this parameter by $\cpw(G)$. 

\begin{theorem}\label{thm:path-fpt}
There is an algorithm that finds a minimum resolving set of a connected graph $G$ in time $f(\bcpw(G))|V(G)|^{\mathcal O(1)}$, where $f$ is a computable function.
\end{theorem}

% We use two properties of graphs that admit path decompositions with small connected bags.
% First, there is a shortest path at bounded distance from every vertex.
% This implies that two sufficiently distant landmarks already distinguish all pairs of vertices that are far apart.
% After guessing these landmarks, we can therefore restrict attention to pairs of nearby vertices and use dynamic programming to resolve them.
% The other case is that all landmarks of an optimum solution lie in a small ball.
% Here we use the second property: a bounded number of nearby vertices separates the ball from the distant part of the graph.
% Distances through this separator allow us to express all remaining requirements by finitely many colors and a monadic second-order formula.
% We can then find an optimum solution by monadic second-order optimization on graphs of bounded treewidth.

\section{Preliminaries}\label{sec:preliminaries}

We write $\dist_G(u,v)$ for the distance between vertices $u,v$ of a connected graph $G$, and omit the subscript when the graph is clear from the context.
For $A\subseteq V(G)$, we write $G[A]$ for the subgraph induced by $A$ and $G-A$ for $G[V(G)\setminus A]$.
The \emph{external vertex boundary} of $A$ is
\[
  \partial_G A=\{v\in V(G)\setminus A: v\text{ has a neighbor in }A\}.
\]
We omit the subscript $G$ when the ambient graph is clear, and use the same notation for a subgraph and its vertex set.

A set $S\subseteq V(G)$ is a \emph{resolving set} if, for every two distinct vertices $u,v$, there is a vertex $s\in S$ with $\dist(s,u)\neq\dist(s,v)$.
A vertex that belongs to a resolving set will also be called a \emph{landmark}.
The \emph{metric dimension} $\mdim(G)$ is the minimum size of a resolving set.
A \emph{leaf} is a vertex of degree one, and its incident edge is a \emph{pendant edge}.
We call a pair of leaves with a common neighbor a \emph{forcing pair}, its members \emph{forcing leaves}, and their common neighbor a \emph{forcing vertex}.
The two leaves have the same distance from every other vertex.
Hence every resolving set contains at least one of them.
We will use such pairs to force a choice of a landmark without prescribing which leaf is chosen.

A \emph{vertex cover} of a graph $F$ is a set $C\subseteq V(F)$ containing at least one endpoint of every edge.
We write $\tau(F)$ for the minimum size of a vertex cover of $F$.
In \VertexCover{}, the input is a graph $F$ and an integer $t$, and the question is whether $\tau(F)\leq t$.

A \emph{tree decomposition} of $G$ consists of a tree $T$ and bags $\beta(t)\subseteq V(G)$, for $t\in V(T)$, such that every edge has both endpoints in some bag and, for every vertex $v$, the nodes whose bags contain $v$ form a nonempty connected subtree of $T$.
Its \emph{width} is $\max_t|\beta(t)|-1$, and the \emph{treewidth} $\tw(G)$ is the minimum possible width.
We call the decomposition \emph{bag-connected} if~$G[\beta(t)]$ is connected for every $t$.
The \emph{bag-connected treewidth} $\ctw(G)$ is the minimum width of a bag-connected tree decomposition of $G$.
In particular, $\tw(G)\leq\ctw(G)$.

We call a rooted tree decomposition \emph{nice} if all bags are nonempty, every leaf has a singleton bag, and every non-leaf node is of one of the following three types.
An \emph{introduce node} $t$ has one child $t'$ and satisfies $\beta(t)=\beta(t')\cup\{v\}$ for some $v\notin\beta(t')$.
A \emph{forget node} $t$ has one child $t'$ and satisfies $\beta(t)=\beta(t')\setminus\{v\}$ for some $v\in\beta(t')$.
A \emph{join node} $t$ has two children $t_1,t_2$ with $\beta(t)=\beta(t_1)=\beta(t_2)$.

A \emph{path decomposition} is a tree decomposition whose underlying tree is a path.
The \emph{bag-connected pathwidth} $\bcpw(G)$ of $G$ is the minimum width over all path decompositions where every bag is required to induce a connected graph.
Thus $\ctw(G)\leq\bcpw(G)$.
The graph-search version of connected pathwidth, denoted by $\cpw(G)$, instead minimizes the width over path decompositions $B_1,\ldots,B_t$ for which $G[\bigcup_{j=1}^i B_j]$ is connected for every $i$.
We write $N_r(v)=\{w\in V(G):\dist(v,w)\leq r\}$ for the radius-$r$ neighborhood, or \emph{ball}, around $v$.

The length of a tree decomposition is the maximum of $\operatorname{diam}_G(\beta(t))$ over its bags, where the diameter is measured in the whole graph.
The tree-length $\tl(G)$ is the minimum length of a tree decomposition~\cite{DourisboureGavoille2007}.
The \emph{induced diameter} of a connected bag is its diameter in the subgraph it induces.
In our hardness construction, every bag has induced diameter at most two.
A cycle is \emph{isometric} if the distance between any two of its vertices is the same in the cycle and in the whole graph.
Every isometric cycle is induced.
We denote the maximum length of an isometric cycle by $\gammaiso(G)$, with value zero for forests.

\section{The tail construction}\label{sec:tails}

In this section, we introduce the graph gadgets used in our reduction, which we call tails, and analyze which vertex pairs remain unresolved by the forced landmarks.

\begin{figure}[ht]
\centering
\begin{tikzpicture}[
  x=1.2cm,y=0.9cm,
  vertex/.style={circle,draw,inner sep=1pt,minimum size=5mm},
  marker/.style={vertex,draw=darkred,fill=darkred!10},
  every edge/.style={draw,thick}]
  \node[vertex] (p) at (0,1) {$p$};
  \node[vertex] (q) at (0,0) {$q$};
  \draw (p)--(q);
  \foreach \i in {1,...,5} {
    \node[vertex] (u\i) at (\i,1) {$u_{\i}$};
    \node[vertex] (v\i) at (\i,0) {$v_{\i}$};
    \draw (u\i)--(v\i);
  }
  \draw (p)--(u1); \draw (q)--(v1);
  \foreach \i [evaluate=\i as \j using int(\i+1)] in {1,...,4} {
    \draw (u\i)--(u\j); \draw (v\i)--(v\j);
  }
  \node[marker] (c) at (3.5,0.5) {$c_4$};
  \draw[darkred] (c)--(u3); \draw[darkred] (c)--(v3);
\end{tikzpicture}
\caption{A ladder with five levels, attached to the edge $pq$, and a marker $c_4$ at level four. The horizontal paths are the rails and the vertical edges $u_hv_h$ are the rungs. The vertices $p,q,c_4$ are outside the ladder.}
\label{fig:tail}
\end{figure}
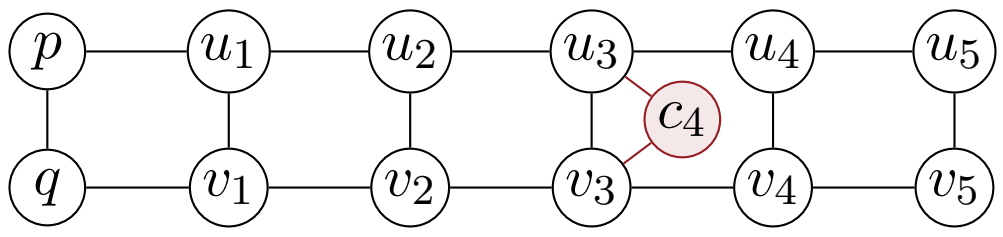

A \emph{ladder with $L$ levels}, where $L\geq2$, consists of two vertex-disjoint paths $u_1,\ldots,u_L$ and $v_1,\ldots,v_L$, together with the edges $u_hv_h$ for $1\leq h\leq L$.
The two paths are its \emph{rails}, their vertices are \emph{rail vertices}, and the edges $u_hv_h$ are its \emph{rungs}.
The vertices $u_h,v_h$ and their rung are at \emph{level} $h$.
The four-cycles $u_{h-1}u_hv_hv_{h-1}u_{h-1}$, for $2\leq h\leq L$, are the \emph{ladder cells}.
We attach the ladder to an edge $pq$ disjoint from its rails by adding $pu_1$ and $qv_1$.
A \emph{marker at level $h$}, for $2\leq h\leq L$, is an additional vertex $c_h$ whose neighbors are $u_{h-1}$ and $v_{h-1}$.
Thus a marker is attached to the rung one level before its own; together with that rung, it forms a \emph{marker triangle}.
\Cref{fig:tail} illustrates these definitions.

For a nonempty set $X=\{x_1,\ldots,x_k\}\subseteq V(G)$ with a fixed ordering, the \emph{distance vector to $X$} of a vertex $v\in V(G)$ is
\[
  \mathbf{d}_X(v)=\bigl(\dist(x_1,v),\ldots,\dist(x_k,v)\bigr).
\]

The following standard lemma expresses distances through a separating set.

\begin{lemma}\label{lem:boundary-transfer}
Let $G$ be a connected graph and let $\varnothing\neq X\subseteq V(G)$.
If $C$ is a component of~$G-X$, $v\in C$, and $s\in V(G)\setminus C$, then
\[
  \dist(s,v)=\min_{x\in X}\bigl(\dist(s,x)+\dist(x,v)\bigr).
\]
In particular, if $v,w$ lie in different components of $G-X$ and satisfy $\mathbf{d}_X(v)=\mathbf{d}_X(w)$, then no vertex outside these two components distinguishes them.
\end{lemma}

% \begin{proof}
% Every $s$-$v$ path meets $X$.
% A shortest path therefore attains one of the terms in the minimum, and the reverse inequality is the triangle inequality.
% The last assertion follows by applying the identity to $v$ and $w$.
% \end{proof}

Our construction attaches several ladders to the same edge $pq$ and adds markers and further local vertices.
A \emph{tail} consists of one ladder and all vertices added locally to it, including its markers and forcing leaves; it excludes $p,q$ and the leaves attached directly to them.
The tails are pairwise disjoint connected subgraphs.
All tails in our construction attach to the same boundary $X=\{p,q\}$.
Thus, writing $H$ for the constructed graph and $T_a$ for its $a$th tail, we have $\partial_H T_a=X$ for every $a$.
We call the forcing leaves within a tail \emph{local forcing leaves}, and the leaves attached directly to $p,q$ \emph{boundary leaves}.
The \emph{entrance graph} of a tail consists of $p,q,u_1,v_1$, the edges $pq,pu_1,u_1v_1,qv_1$, and the paths $pw_pv_1$ and $qw_qu_1$, where $w_p,w_q$ are two additional vertices of the tail.
We specify the remaining local vertices and their edges below.
For the common boundary of our tails, we always use the order $(p,q)$, so
\[
  \mathbf{d}_X(v)=\bigl(\dist(p,v),\dist(q,v)\bigr).
\]
We will repeatedly use \cref{lem:boundary-transfer} to compute distances from vertices outside a tail.
\medskip

The markers in our construction will represent the edges of a \VertexCover{} instance.
For now, we allow an arbitrary choice of the levels at which they occur.
Let $L\geq2$, and let $D_1,\ldots,D_n$ be subsets of $\{2,\ldots,L\}$.
We start with an edge $pq$ and put $X=\{p,q\}$.
For each index $a\in\{1,\ldots,n\}$, we attach a ladder with $L$ levels to $pq$, using the superscript~$a$ to distinguish its vertices from those of the other ladders.
Its rails are $u_1^a,\ldots,u_L^a$ and $v_1^a,\ldots,v_L^a$, its rungs are $u_h^av_h^a$, and its attachment edges are $pu_1^a$ and~$qv_1^a$.
For every $h\in D_a$, we add a marker $c_h^a$ adjacent to $u_{h-1}^a$ and $v_{h-1}^a$.
The marker is thus attached to the rung immediately before its level, as illustrated in \cref{fig:tail}.

\begin{figure}[htbp]
\centering
\begin{subfigure}[b]{0.52\textwidth}
\centering
\resizebox{\linewidth}{!}{%
\begin{tikzpicture}[
  x=1.35cm,y=0.95cm,
  vertex/.style={circle,draw,fill=white,inner sep=1pt,minimum size=6mm},
  landmark/.style={vertex,draw=landmarkblue,fill=landmarkblue!12},
  every edge/.style={draw,thick}]
  % A planar K_4 with two opposite edges subdivided once.
  \node[vertex] (p) at (-3,-1) {$p$};
  \node[vertex] (v) at (3,-1) {$v_1$};
  \node[vertex] (u) at (0,3) {$u_1$};
  \node[vertex] (q) at (0,0) {$q$};
  \node[vertex] (wp) at (0,-1) {$w_p$};
  \node[vertex] (wq) at (0,1.5) {$w_q$};
  \draw (p)--(u)--(v)--(wp)--(p);
  \draw (p)--(q)--(v); \draw (q)--(wq)--(u);
  % Local triangles and their forcing leaves.
  \node[vertex] (sp) at (-1.5,-1.8) {$s_p$};
  \draw (p)--(sp)--(wp);
  \node[landmark] (fp0) at (-2.2,-2.5) {$f_{p,0}$};
  \node[vertex] (fp1) at (-0.8,-2.5) {$f_{p,1}$};
  \draw (sp)--(fp0); \draw (sp)--(fp1);
  \node[vertex] (sq) at (0.35,0.75) {$s_q$};
  \draw (q)--(sq)--(wq);
  \node[landmark] (fq0) at (1.0,0.97) {$f_{q,0}$};
  \node[vertex] (fq1) at (1.2,0.1) {$f_{q,1}$};
  \draw (sq)--(fq0); \draw (sq)--(fq1);
\end{tikzpicture}%
}
\caption{A planar drawing of the entrance and its local forcing vertices.}
\label{fig:core}
\end{subfigure}\hfill
\begin{subfigure}[b]{0.45\textwidth}
\centering
% The first tail of H_{C_4}; the boundary vertices and their leaves are shared.
\begin{tikzpicture}[
  x=1cm,y=1cm,
  every node/.style={font=\footnotesize},
  vertex/.style={circle,draw=black!70,fill=white,inner sep=0pt,
    minimum size=3.3pt,line width=.45pt},
  leaf/.style={vertex,fill=black!8},
  selected/.style={vertex,draw=landmarkblue,fill=landmarkblue,
    minimum size=4.4pt},
  marker/.style={diamond,draw=darkred,fill=darkred!10,inner sep=0pt,
    minimum size=4.4pt,line width=.5pt},
  edge/.style={draw=black!65,line width=.45pt},
  pedge/.style={edge,draw=landmarkblue!85},
  qedge/.style={edge,draw=violet!65!black},
  lab/.style={inner sep=1.5pt},
  every label/.style={font=\footnotesize,inner sep=1.5pt}
]
  % Coordinates keep the boundary attachments outside the local leaves.
  \coordinate (p) at (-2.35,0);
  \coordinate (q) at (2.35,0);
  \coordinate (p0) at (-2.70,.65);
  \coordinate (p1) at (-2.00,.65);
  \coordinate (q0) at (2.00,.65);
  \coordinate (q1) at (2.70,.65);
  \coordinate (sp) at (-1.35,-1.60);
  \coordinate (wp) at (-1.35,-2.35);
  \coordinate (sq) at (1.35,-1.60);
  \coordinate (wq) at (1.35,-2.35);
  \coordinate (fp0) at (-1.70,-.92);
  \coordinate (fp1) at (-1.00,-.92);
  \coordinate (fq0) at (1.00,-.92);
  \coordinate (fq1) at (1.70,-.92);
  \foreach \h in {1,...,6} {
    \coordinate (u\h) at (-.56,{-3.15-.53*(\h-1)});
    \coordinate (v\h) at (.56,{-3.15-.53*(\h-1)});
  }

  \draw[edge,line width=.75pt] (p)--(q);
  \draw[edge] (p0)--(p)--(p1) (q0)--(q)--(q1);
  \foreach \x/\dest/\y in {-1.95/sp/-1.60,-2.10/wp/-2.35,-2.25/u1/-3.15} {
    \draw[pedge,rounded corners=5pt]
      (p) .. controls (-2.35,-.22) and (\x,-.25) .. (\x,-.48)
      --(\x,\y)--(\dest);
  }
  \foreach \x/\dest/\y in {1.95/sq/-1.60,2.10/wq/-2.35,2.25/v1/-3.15} {
    \draw[qedge,rounded corners=5pt]
      (q) .. controls (2.35,-.22) and (\x,-.25) .. (\x,-.48)
      --(\x,\y)--(\dest);
  }
  \draw[edge] (fp0)--(sp)--(fp1) (sp)--(wp);
  \draw[edge] (fq0)--(sq)--(fq1) (sq)--(wq);
  \draw[edge] (wp)--(v1);
  \draw[edge,preaction={draw=white,line width=2.4pt}] (wq)--(u1);
  \foreach \h in {1,...,6} {\draw[edge] (u\h)--(v\h);}
  \foreach \h [evaluate=\h as \j using int(\h+1)] in {1,...,5} {
    \draw[edge] (u\h)--(u\j) (v\h)--(v\j);
  }
  \foreach \h/\r in {2/1,5/4} {
    \coordinate (c\h) at (0,{-3.15-.53*(\r-1)-.265});
    \draw[edge,draw=darkred] (u\r)--(c\h)--(v\r);
    \node[marker] at (c\h) {};
    \node[lab,text=darkred,anchor=west] at (1.18,{-3.15-.53*(\r-1)-.265}) {$c_{\h}^{1}$};
  }

  % Draw vertices after the edges so that every incidence remains visible.
  \node[vertex,label=above:$p$] at (p) {};
  \node[vertex,label=above:$q$] at (q) {};
  \node[selected,label=above:$p^0$] at (p0) {};
  \node[leaf,label=above:$p^1$] at (p1) {};
  \node[selected,label=above:$q^0$] at (q0) {};
  \node[leaf,label=above:$q^1$] at (q1) {};
  \node[vertex,label=right:$s_p^1$] at (sp) {};
  \node[vertex,label=above right:$w_p^1$] at (wp) {};
  \node[vertex,label=left:$s_q^1$] at (sq) {};
  \node[vertex,label=above left:$w_q^1$] at (wq) {};
  \node[selected,label=above:$f_{p,0}^{1}$] at (fp0) {};
  \node[leaf,label=above:$f_{p,1}^{1}$] at (fp1) {};
  \node[selected,label=above:$f_{q,0}^{1}$] at (fq0) {};
  \node[leaf,label=above:$f_{q,1}^{1}$] at (fq1) {};
  \node[selected,label=above left:$u_1^1$] at (u1) {};
  \node[vertex,label=above right:$v_1^1$] at (v1) {};
  \foreach \h in {2,...,6} {
    \node[vertex,label=left:$u_{\h}^{1}$] at (u\h) {};
    \node[vertex,label=right:$v_{\h}^{1}$] at (v\h) {};
  }
\end{tikzpicture}
\caption{A complete tail with $L=6$ and \mbox{$D_1=\{2,5\}$.}}
\label{fig:c4-tail}
\end{subfigure}
\caption{The entrance graph and a complete tail. In (a), deleting $s_p,s_q$ and their leaves gives~$K_4$ with two opposite edges subdivided once. In (b), $p,q$ and their leaves are shared with all other tails. Blue leaves indicate one choice from each depicted forcing pair; the blue vertex $u_1^1$ is an additional landmark used in the reduction. Red diamonds are markers. }
\label{fig:tail-construction}
\end{figure}
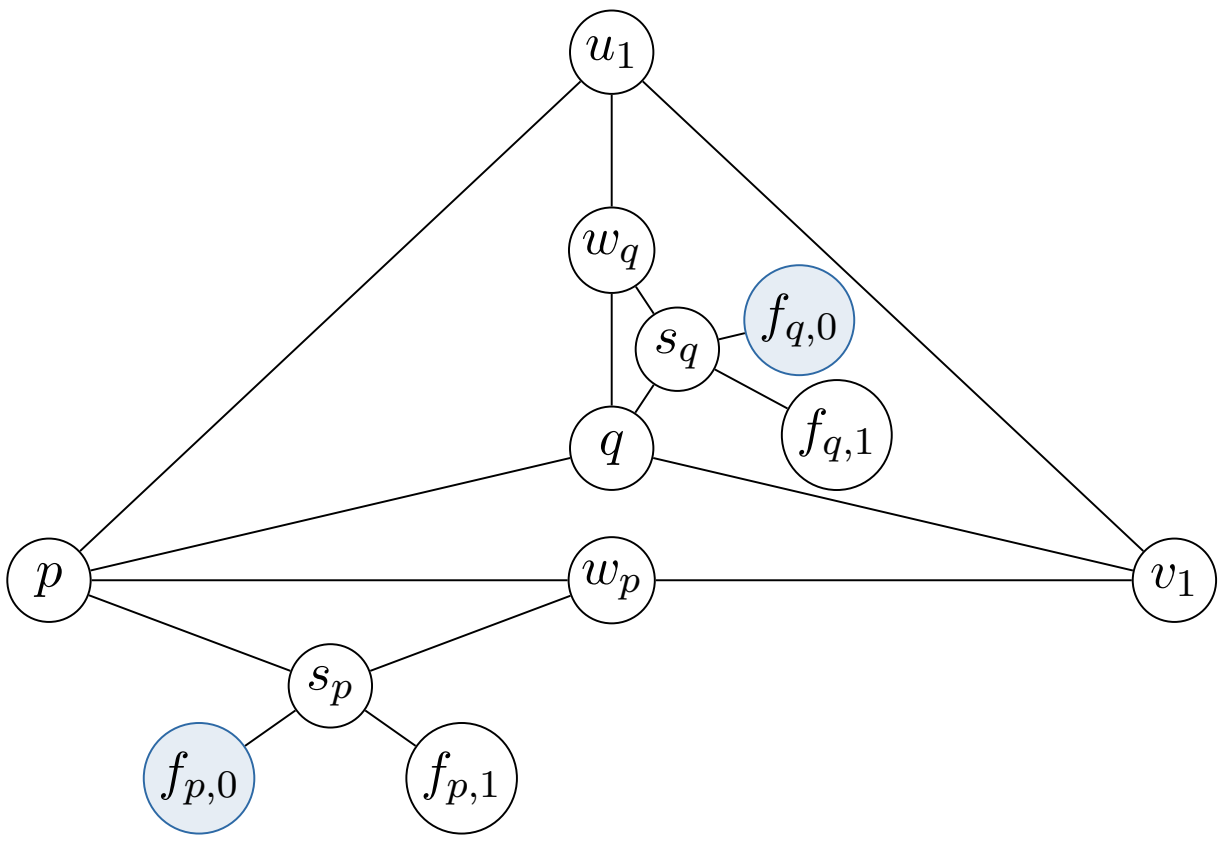
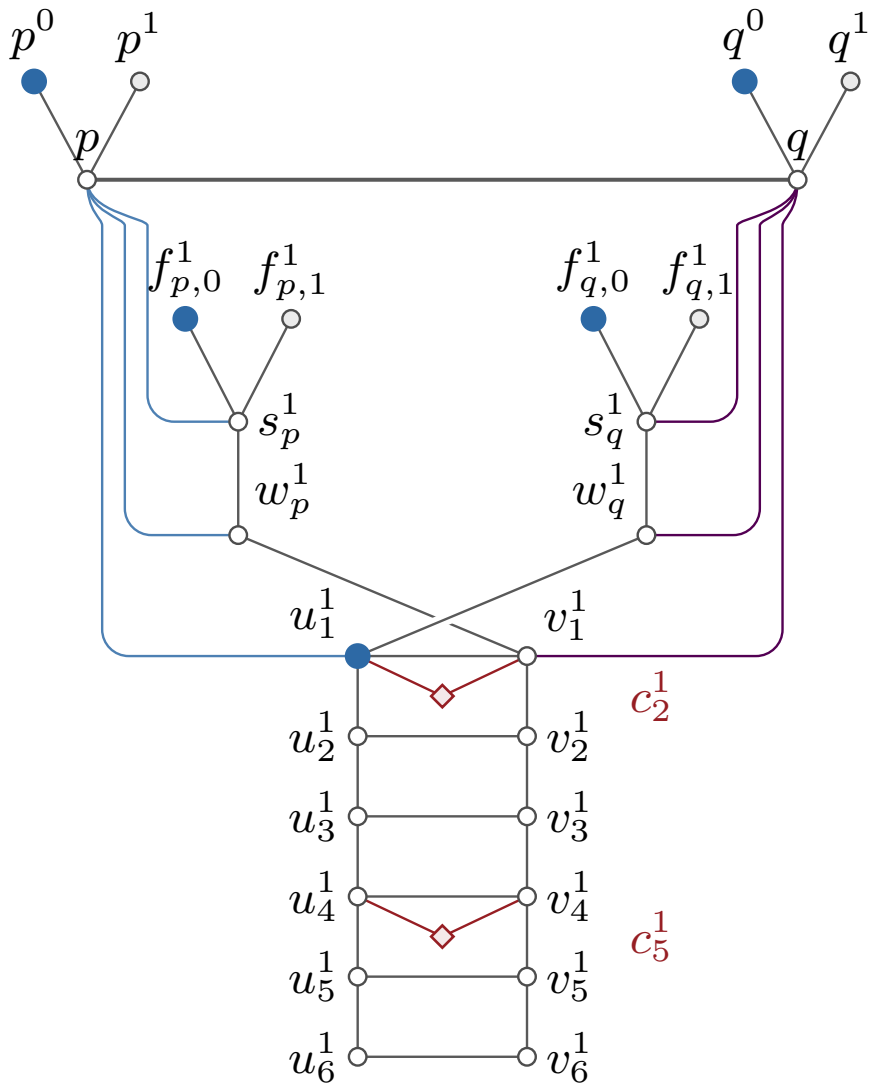

We next arrange that certain landmarks distinguish the rails while leaving markers at the same level unresolved.
For each index $a$ and each $x\in\{p,q\}$, we add vertices $w_x^a,s_x^a$.
We make~$w_p^a$ adjacent to $p,v_1^a$, and $w_q^a$ adjacent to $q,u_1^a$.
The vertices $w_p^a,w_q^a$, together with $p,q,u_1^a,v_1^a$, form the entrance graph of the tail.
We then make $s_x^a$ adjacent to $x,w_x^a$ and attach two leaves $f_{x,0}^a,f_{x,1}^a$ to it.
Thus $s_x^a$ is a local forcing vertex with forcing pair $\{f_{x,0}^a,f_{x,1}^a\}$.
Finally, we attach two leaves~$p^0,p^1$ to $p$ and two leaves $q^0,q^1$ to $q$.
There are no further edges, and we denote the resulting graph by $H$.
The tail $T_a$ consists of all vertices introduced for index $a$, including its local forcing leaves but excluding~$p,q$ and their leaves.
Each tail is a component of $H-X$ and has external boundary $\partial_H T_a=X$.
\Cref{fig:tail-construction} shows the entrance graph and a complete example of a tail.

To see the role of these vertices, observe that $s_p^a$ has distance two from both $u_1^a$ and~$v_1^a$, although it is closer to $p$ than to $q$.
It therefore distinguishes the $v$-rail of its own tail from the $v$-rail of another tail.
The vertex $s_q^a$ does the same for the $u$-rails.
Neither vertex distinguishes markers at the same level.
Their leaves inherit these distance properties, and every resolving set must contain at least one leaf from each pair.

\begin{lemma}\label{lem:distances}
The distance vectors $\mathbf{d}_X(v)$ of the vertices in $H$ are as follows, with coordinates ordered as $(p,q)$.
\begin{center}
\begin{tabular}{@{}lc@{}}
\toprule
Vertex $v$ & Distance vector $\mathbf{d}_X(v)$ \\
\midrule
$p$, $q$ & $(0,1)$, $(1,0)$ \\
$p^j$, $q^j$ & $(1,2)$, $(2,1)$ \\
$u_h^a$, $v_h^a$ & $(h,h+1)$, $(h+1,h)$ \\
$c_h^a$ & $(h,h)$ \\
$s_p^a$, $w_p^a$ & $(1,2)$ \\
$s_q^a$, $w_q^a$ & $(2,1)$ \\
$f_{p,j}^a$, $f_{q,j}^a$ & $(2,3)$, $(3,2)$ \\
\bottomrule
\end{tabular}
\end{center}
For every local forcing vertex $s_x^a$ and every marker $c_h^b$, regardless of whether $a=b$, we have
\begin{equation}\label{eq:neutral-markers}
  \dist(s_x^a,c_h^b)=h+1,
  \qquad
  \dist(f_{x,j}^a,c_h^b)=h+2.
\end{equation}
\end{lemma}

\begin{proof}
We first compute the distances to the boundary.
A shortest path cannot pass through a marker as an internal vertex, since the two neighbors of the marker are joined by a rung.
We may therefore ignore the markers when computing distances between the other vertices.
Along the ladder, $u_h^a$ has distance $h$ from $p$ and needs one extra step to reach $q$; the distances for $v_h^a$ are symmetric.
The paths through $s_x^a,w_x^a$ give no shorter route to either end of the first rung, and hence do not change these distances.
As $c_h^a$ is adjacent to both rail vertices at level $h-1$, its distance from each boundary vertex is $h$.
We read off the remaining entries from the beginning of the tail, adding one step when we pass to a forcing leaf.

We next verify \cref{eq:neutral-markers}.
In its own tail, $s_x^a$ has distance two from both first rail vertices, and hence distance $h$ from both rail vertices at level $h-1$.
Its distance to $c_h^a$ is therefore $h+1$.
For a marker in another tail, we obtain the same value from \cref{lem:boundary-transfer}: $s_x^a$ has distance one from the boundary, and $\mathbf{d}_X(c_h^b)=(h,h)$.
Again, passing to a forcing leaf adds one step.
\end{proof}

We now show that the forced landmarks leave precisely the marker pairs that we want to use in the reduction.

\begin{lemma}\label{lem:forced-collisions}
Choose one leaf from each of the $2n+2$ forcing pairs, and let $S_0$ be the resulting set.
The pairs not resolved by $S_0$ are precisely the pairs of markers at the same level in different tails.
\end{lemma}

\begin{proof}
A selected vertex distinguishes itself from every other vertex, so we consider only pairs outside $S_0$.
For each $x\in\{p,q\}$, let $\widehat x$ be the selected leaf adjacent to $x$.
For every $y\neq\widehat x$, we have $\dist(\widehat x,y)=1+\dist(x,y)$.
The two selected boundary leaves therefore distinguish any two vertices outside $S_0$ with different distance vectors to $X$.
We are left with pairs having the same distance vector to $X$, which we examine using \cref{lem:distances}.

Let $\widehat f_x^a$ be the selected leaf at $s_x^a$.
Its distances to $s_x^a$ and $w_x^a$ are one and two, respectively.
Every other vertex outside $S_0$ with their common distance vector to $X$ is at distance at least three from $\widehat f_x^a$.
Thus all pairs involving an $s$-vertex or a $w$-vertex are resolved.
Similarly, the unselected leaf at $s_x^a$ has distance two from the selected leaf of its forcing pair, while every other vertex outside~$S_0$ with the same distance vector to $X$ has distance at least four.
This also resolves all pairs involving a local forcing leaf.

A rail vertex $u_h^a$ can share its distance vector to $X$ with a remaining vertex only if that vertex is $u_h^b$ in another tail or, when $h=1$, the unselected leaf at $p$.
The selected leaf $\widehat f_q^a$ has distance~$h+2$ from $u_h^a$ and distance $h+3$ from each of these other vertices, so it distinguishes them.
We argue symmetrically with $\widehat f_p^a$ for the vertices $v_h^a$.
The boundary vertices and the two remaining boundary leaves have distinct distance vectors to $X$ from one another.
The only pairs left are therefore markers at the same level in different tails.

These marker pairs are indeed unresolved: the selected boundary leaves have distance~$h+1$ from every marker at level $h$, and all selected local leaves have distance $h+2$ by \cref{eq:neutral-markers}.
\end{proof}

\section{The reduction}

We reduce from \VertexCover{}, which is NP-complete~\cite[Problem~GT1]{GareyJohnson1979}.
Let $F$ be an input graph with $n$ vertices and $m$ edges, and let $t$ be the requested cover size.
We assign every edge $e$ a distinct level $\lambda(e)\in\{2,\ldots,m+1\}$.
For each vertex $a\in V(F)$, let $D_a$ be the set of levels assigned to edges incident with $a$.
We apply the construction of \cref{sec:tails} with these sets and with $L=m+2$, denote the resulting graph by $H_F$, and set $K=2n+2+t$.
Each edge $ab$ thus gives two markers at its level, one in each endpoint tail.

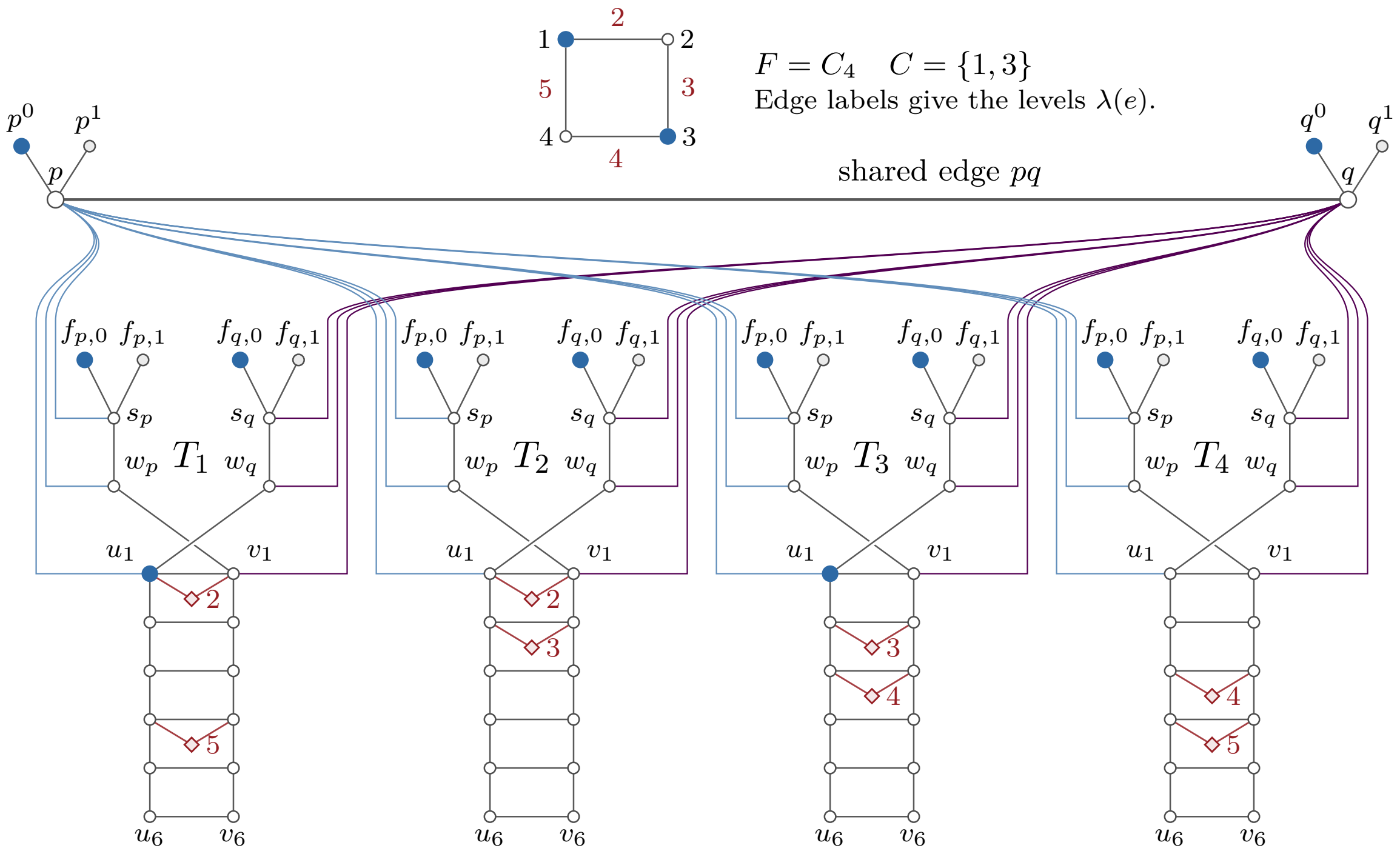
\begin{figure}[!htb]
\centering
\resizebox{\textwidth}{!}{% Complete reduction of C_4. Edge levels in cyclic order: 2,3,4,5.
% The coordinates describe a schematic drawing, not a planar embedding.
\begin{tikzpicture}[
  x=1cm,y=1cm,
  font=\scriptsize,
  vertex/.style={circle,draw=black!70,fill=white,inner sep=0pt,
    minimum size=3.3pt,line width=.45pt},
  leaf/.style={vertex,fill=black!8},
  selected/.style={vertex,draw=landmarkblue,fill=landmarkblue,
    minimum size=4.4pt},
  marker/.style={diamond,draw=darkred,fill=darkred!10,inner sep=0pt,
    minimum size=4.4pt,line width=.5pt},
  edge/.style={draw=black!65,line width=.45pt},
  p edge/.style={draw=landmarkblue!75,line width=.4pt},
  q edge/.style={draw=violet!65!black,line width=.4pt},
  marker edge/.style={draw=darkred!85,line width=.5pt},
  every label/.append style={inner sep=1.5pt}
]
  % The small input graph uses the same selection and edge-level colors.
  \node[selected,label=left:{$1$}] (a1) at (5.85,1.65) {};
  \node[vertex,label=right:{$2$}] (a2) at (6.9,1.65) {};
  \node[selected,label=right:{$3$}] (a3) at (6.9,.65) {};
  \node[vertex,label=left:{$4$}] (a4) at (5.85,.65) {};
  \draw[edge] (a1)--node[above,text=darkred] {$2$} (a2)
    --node[right,text=darkred] {$3$} (a3)
    --node[below,text=darkred] {$4$} (a4)
    --node[left,text=darkred] {$5$} (a1);
  \node[anchor=west,align=left,font=\footnotesize] at (7.65,1.2)
    {$F=C_4$\quad $C=\{1,3\}$\\[-1pt]
     \scriptsize Edge labels give the levels $\lambda(e)$.};

  \coordinate (p) at (.6,0);
  \coordinate (q) at (13.9,0);
  \draw[edge,line width=.8pt] (p)--(q);
  \node[above,font=\footnotesize] at (9.7,0) {shared edge $pq$};
  \coordinate (p0) at (.25,.55);
  \coordinate (p1) at (.95,.55);
  \coordinate (q0) at (13.55,.55);
  \coordinate (q1) at (14.25,.55);
  \draw[edge] (p0)--(p)--(p1) (q0)--(q)--(q1);

  % All connections to the common boundary are explicit; intersections
  % without a node are only crossings. The three routes on each side
  % end at s_x, w_x, and the first rail vertex, respectively.
  \foreach \a/\cx in {1/2,2/5.5,3/9,4/12.5} {
    \coordinate (sp\a) at (\cx-.80,-2.25);
    \coordinate (wp\a) at (\cx-.80,-2.95);
    \coordinate (sq\a) at (\cx+.80,-2.25);
    \coordinate (wq\a) at (\cx+.80,-2.95);
    \foreach \h in {1,...,6} {
      \coordinate (u\a-\h) at (\cx-.43,{-3.85-.5*(\h-1)});
      \coordinate (v\a-\h) at (\cx+.43,{-3.85-.5*(\h-1)});
    }
    \draw[p edge] (p) .. controls +(1,-.6) and +(0,.6) ..
      (\cx-1.4,-1.25) -- (\cx-1.4,-2.25) -- (sp\a);
    \draw[p edge] (p) .. controls +(1,-.6) and +(0,.6) ..
      (\cx-1.5,-1.25) -- (\cx-1.5,-2.95) -- (wp\a);
    \draw[p edge] (p) .. controls +(1,-.6) and +(0,.6) ..
      (\cx-1.6,-1.25) -- (\cx-1.6,-3.85) -- (u\a-1);
    \draw[q edge] (q) .. controls +(-1,-.6) and +(0,.6) ..
      (\cx+1.4,-1.25) -- (\cx+1.4,-2.25) -- (sq\a);
    \draw[q edge] (q) .. controls +(-1,-.6) and +(0,.6) ..
      (\cx+1.5,-1.25) -- (\cx+1.5,-2.95) -- (wq\a);
    \draw[q edge] (q) .. controls +(-1,-.6) and +(0,.6) ..
      (\cx+1.6,-1.25) -- (\cx+1.6,-3.85) -- (v\a-1);
    \draw[edge] (sp\a)--(wp\a)--(v\a-1);
    \draw[edge,preaction={draw=white,line width=1.8pt}]
      (sq\a)--(wq\a)--(u\a-1);
    \foreach \x/\dx in {p/-.80,q/.80} {
      \coordinate (f\x\a0) at (\cx+\dx-.3,-1.65);
      \coordinate (f\x\a1) at (\cx+\dx+.3,-1.65);
      \draw[edge] (f\x\a0)--(s\x\a)--(f\x\a1);
    }
    \foreach \h in {1,...,6} {
      \draw[edge] (u\a-\h)--(v\a-\h);
    }
    \foreach \h [evaluate=\h as \j using int(\h+1)] in {1,...,5} {
      \draw[edge] (u\a-\h)--(u\a-\j) (v\a-\h)--(v\a-\j);
    }
  }
  % Two markers for each input edge, attached one rung before its level.
  \foreach \a/\cx/\h in {1/2/2,1/2/5,2/5.5/2,2/5.5/3,
                           3/9/3,3/9/4,4/12.5/4,4/12.5/5} {
    \pgfmathtruncatemacro{\rung}{\h-1}
    \coordinate (c\a-\h) at (\cx,{-3.85-.5*(\h-2)-.26});
    \draw[marker edge] (u\a-\rung)--(c\a-\h)--(v\a-\rung);
    \node[marker,label={[text=darkred]right:{$\h$}}] at (c\a-\h) {};
  }
  % Paint the vertices above all edges so incidence is unambiguous.
  \foreach \a/\cx in {1/2,2/5.5,3/9,4/12.5} {
    \node[font=\normalsize] at (\cx,-2.65) {$T_{\a}$};
    \node[vertex,label=right:{$s_p$}] at (sp\a) {};
    \node[vertex,label=left:{$s_q$}] at (sq\a) {};
    \node[vertex,label=above right:{$w_p$}] at (wp\a) {};
    \node[vertex,label=above left:{$w_q$}] at (wq\a) {};
    \foreach \x in {p,q} {
      \node[selected,label=above:{$f_{\x,0}$}] at (f\x\a0) {};
      \node[leaf,label=above:{$f_{\x,1}$}] at (f\x\a1) {};
    }
    \foreach \h in {1,...,6} {
      \node[vertex] at (u\a-\h) {};
      \node[vertex] at (v\a-\h) {};
    }
    \node[above left] at (u\a-1) {$u_1$};
    \node[above right] at (v\a-1) {$v_1$};
    \node[below] at (u\a-6) {$u_6$};
    \node[below] at (v\a-6) {$v_6$};
  }
  \foreach \a in {1,3} {\node[selected] at (u\a-1) {};}
  \node[vertex,minimum size=4.8pt,label=above:{$p$}] at (p) {};
  \node[vertex,minimum size=4.8pt,label=above:{$q$}] at (q) {};
  \node[selected,label=above:{$p^0$}] at (p0) {};
  \node[leaf,label=above:{$p^1$}] at (p1) {};
  \node[selected,label=above:{$q^0$}] at (q0) {};
  \node[leaf,label=above:{$q^1$}] at (q1) {};
\end{tikzpicture}}
\caption{The graph $H_{C_4}$ and the resolving set arising from $C=\{1,3\}$, shown in blue. Red diamonds are markers labeled by their levels; the two markers at each level represent one input edge. Superscripts identifying the tails are omitted within the drawing. Blue and purple attachment edges lead to $p$ and $q$, respectively. }
\label{fig:c4-reduction}
\end{figure}

\Cref{fig:c4-reduction} illustrates the reduction for the cycle $F=C_4$ on vertices $1,2,3,4$, with edge levels $\lambda(12)=2$, $\lambda(23)=3$, $\lambda(34)=4$, and $\lambda(41)=5$.
Here $L=6$, and the first tail is the one shown in \cref{fig:c4-tail}.
The vertex cover $C=\{1,3\}$ gives the displayed resolving set: one leaf from each of the ten forcing pairs, together with $u_1^1$ and $u_1^3$.

The boundary and its four leaves contribute six vertices.
Each tail has $2L$ rail vertices, eight additional vertices at its beginning, and one marker for every incident input edge.
Since every input edge contributes two markers, the total number of vertices is
\[
  |V(H_F)|=6+n(2L+8)+2m=6+2n(m+6)+2m.
\]
We can construct the graph and the decomposition described below in $\mathcal O(n(m+1))$ time.

The following identity proves the correctness of the reduction.
Recall that $\mdim(H_F)$ is the minimum size of a resolving set of $H_F$, and $\tau(F)$ is the minimum size of a vertex cover of $F$.

\begin{theorem}\label{thm:exact-dimension}
For every input graph $F$ on $n$ vertices,
\[
  \mdim(H_F)=2n+2+\tau(F).
\]
In particular, $F$ has a vertex cover of size at most $t$ if and only if $\mdim(H_F)\leq K$.
\end{theorem}

\begin{proof}
We first extract a vertex cover from a resolving set $S$ of $H_F$.
It contains at least one leaf from each forcing pair, and hence at least $2n+2$ forcing leaves.
For an edge $ab$ of $F$, consider its two markers $c_h^a,c_h^b$, where $h=\lambda(ab)$.
Their distance vectors to $X$ agree, so no vertex outside $T_a\cup T_b$ distinguishes them by \cref{lem:boundary-transfer}.
Nor does any forcing leaf distinguish them, by \cref{eq:neutral-markers} and the distances from the boundary leaves.
Consequently, $S$ must contain a vertex of $T_a\cup T_b$ that is not a forcing leaf.
We take all vertices $a\in V(F)$ whose tails contain such a landmark.
They form a vertex cover of size at most $|S|-(2n+2)$, since the tails are disjoint.
It follows that $|S|\geq2n+2+\tau(F)$.

Conversely, let $C$ be a minimum vertex cover of $F$.
We choose one leaf from each forcing pair and add $u_1^a$ for every $a\in C$.
By \cref{lem:forced-collisions}, we only need to resolve pairs of markers at the same level.
Each such pair represents an edge $ab$ of $F$, and at least one endpoint, say $a$, belongs to $C$.
Following the $u$-rail inside $T_a$, we obtain
\[
  \dist(u_1^a,c_h^a)\leq h-1<h+1
  =\dist(u_1^a,c_h^b),
\]
where the last equality follows from \cref{lem:boundary-transfer}.
Thus the additional landmarks resolve every remaining pair.
We have chosen $2n+2+\tau(F)$ landmarks in total, which proves the reverse inequality.
\end{proof}

\section{Structural properties of the construction}\label{sec:structure}

We now verify the structural properties required in \cref{thm:main}.
The main point is to understand the graph at the beginning of a tail, the ladders and forcing leaves can then be attached along edges or single vertices.

\begin{lemma}\label{lem:structure}
For every input graph $F$, $H_F$ is planar and satisfies
\[
  \tw(H_F)=\ctw(H_F)=3,\qquad
  \tl(H_F)=2,\qquad
  \gammaiso(H_F)=4.
\]
Every induced cycle has length at most four.
A bag-connected tree decomposition of width three whose bags have induced diameter at most two can be constructed in $\mathcal O(|V(H_F)|)$ time.
\end{lemma}

\begin{proof}
We construct a bag-connected tree decomposition as follows.
We take $X=\{p,q\}$ as the root bag and attach the bags $\{p,p^0,p^1\}$ and $\{q,q^0,q^1\}$ to it.
For each tail $T_a$, we attach the bag $I_a=\{p,q,u_1^a,v_1^a\}$ to the root.
From $I_a$, we cover the local forcing vertices and their leaves by the two paths of bags
\[
 I_a
 -\{x,u_1^a,v_1^a,w_x^a\}
 -\{x,w_x^a,s_x^a\}
 -\{s_x^a,f_{x,0}^a,f_{x,1}^a\},
 \qquad x\in\{p,q\}.
\]
We cover the ladder by a further path from $I_a$, namely
\[
 I_a-P_2^a-Q_2^a-\cdots-P_L^a-Q_L^a,
\]
where
\[
 P_h^a=\{u_{h-1}^a,v_{h-1}^a,u_h^a\},\qquad
 Q_h^a=\{v_{h-1}^a,u_h^a,v_h^a\}.
\]
Whenever $h\in D_a$, we attach the marker bag $\{u_{h-1}^a,v_{h-1}^a,c_h^a\}$ at $P_h^a$.

Every edge is covered, and the bags containing any fixed vertex form a subtree: boundary vertices occur through the root, rail vertices in consecutive ladder bags and their incident attachments, and the remaining vertices along their local paths of bags.
Every bag has at most four vertices and induces a connected graph of diameter at most two, proving $\ctw(H_F)\leq3$ and $\tl(H_F)\leq2$.
To obtain the reverse bound on treewidth, we contract $pw_p^a$ and $qw_q^a$ at the beginning of any tail.
The two contracted vertices together with $u_1^a,v_1^a$ form a $K_4$.
Hence $3\leq\tw(H_F)\leq\ctw(H_F)\leq3$.

For planarity and the bound on induced cycles, we consider the entrance graph on $p,q,u_1,v_1,w_p,w_q$, omitting the superscript $a$.
As shown in \cref{fig:core}, it is obtained from~$K_4$ on $p,q,u_1,v_1$ by subdividing the two opposite edges $pv_1$ and $qu_1$ once.
In particular, it is planar.
We obtain the whole graph by gluing entrance graphs along~$pq$, ladder cells and marker triangles along rung edges, and the triangles $xw_xs_x$ along $xw_x$, before adding the pendant edges.
Each piece is planar, and gluing along an edge or a vertex preserves planarity.
Thus $H_F$ is planar.

The entrance graph is bipartite, with parts $\{p,v_1,w_q\}$ and $\{q,u_1,w_p\}$, and has six vertices and eight edges.
An induced cycle longer than four would therefore have to use all six vertices, but the remaining two edges would then be chords.
Hence every induced cycle in the entrance graph has length at most four.
An induced cycle of length at least four cannot cross a clique along which two graphs are glued: two vertices of that clique would give a chord, or a single vertex would disconnect the cycle.
Every induced cycle of $H_F$ therefore lies in an entrance graph, a ladder cell, or one of the attached triangles.
This proves the bound of four and hence $\gammaiso(H_F)\leq4$.
A ladder cell is an induced, and therefore isometric, four-cycle, so equality holds.
Finally, a graph has tree-length at most one exactly when it is chordal, whereas $H_F$ contains an induced four-cycle.
Together with the decomposition above, this gives $\tl(H_F)=2$.
\end{proof}

\begin{lemma}\label{lem:pathwidth-four}
Every graph in the tail construction has a path decomposition of width at
most four. The decomposition can be constructed in linear time and the
union of every initial sequence of bags induces a connected graph.
\end{lemma}

\begin{proof}
We keep $p$ and $q$ in every bag and process the tails one after another.
We start with~$\{p,q\}$, followed by one bag for each of the four boundary
leaves, containing that leaf together with $p,q$.

For a tail $a$, we use the following sequence of sets, adding $p,q$ to
each of them. We omit the superscript $a$ to keep the notation short.
\[
\begin{gathered}
 \{s_p,f_{p,0}\},\quad
 \{s_p,f_{p,1}\},\quad
 \{s_p,w_p\},\quad
 \{w_p,v_1\},\\
 \{v_1,s_q,f_{q,0}\},\quad
 \{v_1,s_q,f_{q,1}\},\quad
 \{v_1,s_q,w_q\},\quad
 \{v_1,w_q,u_1\}.
\end{gathered}
\]
These bags cover the edges at the beginning of the tail, and the last
one contains both ends of the first rung. For $h=2,\ldots,L$, we continue
with the sets
\[
 \{u_{h-1},v_{h-1},u_h\},\qquad
 \{v_{h-1},u_h,v_h\},
\]
again adding $p,q$. If $h\in D_a$, we insert the marker bag immediately
before these two bags:
\[
 \{p,q,u_{h-1},v_{h-1},c_h\}.
\]

Every edge is covered, every vertex appears in consecutive bags, and
every bag has at most five vertices. After the first bag, each new vertex
has a path to an earlier vertex using only vertices of the current bag.
Thus every initial union of bags induces a connected graph.
The construction uses linearly many bags and takes linear time.
\end{proof}

% \todo[inline]{The following paragraph should be clearly explained. If a graph with small geodesic cycles have a tree-decomposition of small width, it can be made connected with small width. One would be quick to think that the same should be true for path-decompositions, but that might not be the case. I think it should be clearly explained that the constructed graph may have large bag-connected pathwidth. The following paragraph was not so clear to me.}

% Proposed replacement addressing the preceding TODO; blue for author review.
\medskip\noindent \textbf{Remark:}
\begingroup\color{black}
For tree decompositions, bounded treewidth together with a bound on the length of isometric cycles implies bounded bag-connected treewidth~\cite{DiestelMuller2018,HamannWeissauer2016}.
The analogous statement for path decompositions is false.
In \cref{lem:pathwidth-four}, the union of each initial sequence of bags is connected, but an individual bag need not be: a bag containing $p,q$ and rail vertices far along a tail omits the intervening rail vertices.
In fact, our graphs have prefix-connected pathwidth at most four and no induced cycle longer than four, but their bag-connected pathwidth is unbounded.
We prove that $\bcpw(H_F)\geq L$ whenever $F$ has at least three vertices.

Consider any path decomposition $B_1,\ldots,B_t$ with connected bags.
Choose three tails and a bag containing the last rail vertex of each tail; relabel these tails so that $u_L^a\in B_{i_a}$ for $a\in\{1,2,3\}$ and $i_1\leq i_2\leq i_3$.
Let $P$ be the path from $u_L^1$ along its rail to $p$ and then along the third rail to $u_L^3$.
The bags meeting a connected subgraph form an interval, since the occurrence intervals of adjacent vertices overlap.
The bags meeting $P$ therefore form an interval containing $i_1$ and $i_3$, so $B_{i_2}$ contains some vertex $x\in V(P)$ as well as $u_L^2$.
Since $P$ lies outside $T_2$, every $u_L^2$--$x$ path meets $\{p,q\}$.
By $\mathbf{d}_X(u_L^2)=(L,L+1)$, these two vertices have distance at least $L$.
The connected graph $H_F[B_{i_2}]$ must contain a path between them of length at least $L$, and hence $|B_{i_2}|\geq L+1$.
This proves the lower bound.
Since $L=|E(F)|+2$ can be arbitrarily large, the width-four decomposition with connected initial unions cannot in general be replaced by one of bounded width with connected bags.
\par\endgroup

\begin{proof}[Proof of \cref{thm:main}]
The construction is polynomial, and \cref{thm:exact-dimension} proves that it is a correct reduction from \VertexCover{}.
The required structural properties and the two decompositions follow from \cref{lem:structure,lem:pathwidth-four}.
Finally, we can verify a proposed resolving set in polynomial time by computing its distance vectors, which proves membership in NP.
\end{proof}

\section{Path decompositions with connected bags}\label{sec:path-algorithm}

We now prove \cref{thm:path-fpt}, which states that a minimum resolving set of a connected graph~$G$ can be found in time $f(\bcpw(G))|V(G)|^{\mathcal O(1)}$ for some computable function $f$.
Our proof uses the fact that a graph with a path decomposition of small bag diameter lies close to a shortest path.
We distinguish two cases according to whether a minimum resolving set contains two distant vertices or is contained in a small ball.
The algorithm will use a dominating shortest path and small separators, which can be found without computing a path decomposition with connected bags.
Throughout this section, distances and bag diameters are measured in the whole graph.

\subsection{Distant landmarks}

A path is \emph{$q$-dominating} in a graph $G$ if every vertex of $G$ is at distance at most $q$ from it.

\begin{lemma}\label{lem:find-dominating-geodesic}
If a connected graph has a path decomposition whose bags have diameter at most $q$, then it has a $q$-dominating shortest path, which can be found in polynomial time.
\end{lemma}

\begin{proof}
We compute an arbitrary shortest path between every pair of vertices and also consider one-vertex paths.
Among these paths, we choose one minimizing the maximum distance of a vertex from the path.

To see that one of these paths is $q$-dominating, consider a path decomposition $B_1,\ldots,B_t$ with bags of diameter at most $q$.
After discarding empty bags, choose $a\in B_1$ and $b\in B_t$.
The bags meeting a connected subgraph form an interval.
Every $a$-$b$ path therefore meets every bag, since this interval contains both $1$ and $t$.
In particular, the computed shortest path for $a,b$ meets every bag.
Every vertex shares a bag with a vertex of this path, so the path is $q$-dominating.
\end{proof}

\begin{lemma}\label{lem:distant-landmarks}
Suppose $G$ has a $q$-dominating shortest path.
If $\dist(s,t)>6q$, then $\{s,t\}$ distinguishes every pair $u,v$ with $\dist(u,v)>6q$.
\end{lemma}

\begin{proof}
Index the vertices of the shortest path by their distance from one endpoint.
For every vertex $x$, choose a path vertex within distance $q$ and denote its index by $z_x$.
Because the path is shortest, the triangle inequality gives
\[
 \bigl|\dist(x,y)-|z_x-z_y|\bigr|\leq2q.
\]
Suppose $u,v$ are not distinguished by $s$, and order their projections so that $z_u<z_v$.
Their distance is greater than $6q$, so $z_v-z_u>4q$.
Equality of the two distances from $s$ implies
\[
 \bigl||z_s-z_u|-|z_s-z_v|\bigr|\leq4q.
\]
Thus $z_s$ lies in the interval of length $4q$ centered at $(z_u+z_v)/2$.
If $t$ also fails to distinguish~$u,v$, its index lies in the same interval.
It follows that $\dist(s,t)\leq |z_s-z_t|+2q\leq6q$, a contradiction.
\end{proof}

It follows that, once we have guessed two landmarks at distance greater than $6q$, only pairs at distance at most $6q$ need further attention.
If a minimum resolving set contains no such two landmarks, then it is contained in the ball of radius $6q$ around any of its vertices.
We next show how to find a smallest resolving set in each of these two cases.

\subsection{Resolving nearby pairs}

For pairs at bounded distance, we can work with a tree decomposition.
We need both its width and its bag diameter to be bounded.
The following approximation result shows that such a decomposition can be found efficiently.

\begin{lemma}[\cite{ChakrabortyEtAlConnectedTreeDecompositions}]\label{lem:short-tree-decomposition}
Given a graph $G$ with $n$ vertices and $m$ edges, we can compute a bag-connected tree decomposition of width $\mathcal O(c^2)$ in time $2^{\mathcal O(c^2)}n^{\mathcal O(1)}$, where $c:=\ctw(G)$.
\end{lemma}

A connected bag of size at most $w+1$ has diameter at most $w$, so this also bounds the bag diameter by $\mathcal O(c^2)$.

\begin{lemma}\label{lem:local-resolving-dp}
Let $r,w,\ell$ be nonnegative integers. Given a connected graph $G$ and a
tree-decomposition $(T,\beta)$ of width at most $w$ whose bags have diameter at most
$\ell$ in $G$, one can find a smallest set distinguishing all pairs of
distinct vertices at distance at most $r$ in time
$f(w,\ell,r)(|V(G)|+|V(T)|)^{\mathcal O(1)}$. The same holds if specified vertices are
required to belong to the set or forbidden from belonging to it.
\end{lemma}

\begin{proof}
We first compute all distances in $G$.
These are the distances used in the dynamic program, even when both vertices lie in a part of the graph that has already been processed.
We refine the given decomposition into a nice rooted decomposition.
Every intermediate bag is a subset of an original bag, so the bounds on width and diameter are preserved.
Since $G$ is connected, adjacent nonredundant bags have nonempty intersection.
Refining through these intersections, we may therefore assume that all bags are nonempty and that the leaves have singleton bags.

Consider a node with bag $X$, and let $V_X$ be the union of the bags in its subtree.
We need to remember which pairs at distance at most $r$ inside $V_X$ still require a landmark outside, and which vertices inside $V_X$ may form such a pair with a vertex outside.
The latter vertices must lie within distance $r$ of $X$, because $X$ separates the two parts of the graph.
For a vertex~$z$, write
\[
 a_X(z)=\min_{x\in X}\dist(z,x),\qquad
 p_X(z)=\bigl(\dist(z,x)-a_X(z):x\in X\bigr).
\]
We call $p_X(z)$ the profile of $z$ at $X$.
Its coordinates belong to $\{0,\ldots,\ell\}$.
For a partial solution $S\subseteq V_X$, we remember $S\cap X$ and the following three sets.

The set $P$ consists of the profiles of vertices in $S$.
It describes how these landmarks distinguish vertices on the other side of $X$.
For each pair $u,v\in V_X$ with $0<\dist(u,v)\le r$ that is not yet distinguished by $S$, we keep its two distance vectors to $X$, after subtracting the smallest coordinate occurring in either vector.
Let $D$ be the set of these records. Since
$|a_X(u)-a_X(v)|\le r$, every coordinate of a record belongs to
$\{0,\ldots,r+\ell\}$.

To deal with pairs having one vertex on either side of the bag, for each $u\in V_X$ with $a_X(u)\le r$ we keep
\[
 \alpha_u=\mathbf{d}_X(u),\qquad
 \sigma_u=\{(p_X(s),\dist(s,u)-a_X(s)):s\in S\}.
\]
Let $C$ be the set of these pairs $(\alpha_u,\sigma_u)$.
The vector $\alpha_u$ gives the distances of $u$ to the bag, while $\sigma_u$ describes its distances to the chosen landmarks relative to their distances from the bag.
Every coordinate of $\alpha_u$ is at most $r+\ell$, and the triangle inequality gives
\[
 -r\le \dist(s,u)-a_X(s)\le r+\ell.
\]
The number of possible records therefore depends only on $w,\ell,r$.
We keep them as sets, without multiplicities.
For each choice of $S\cap X,P,D,C$, the table stores the smallest cardinality of a partial solution with this information.

We verify that this information is sufficient.
Suppose first that a vertex $v$ is introduced.
For every previously processed vertex $z$, separation by the old bag $X$ gives
\[
 \dist(z,v)=\min_{x\in X}\bigl(\dist(z,x)+\dist(x,v)\bigr).
\]
Thus the old coordinates determine the new coordinate at $v$ in every recorded vector.
They also show that the distance of an old vertex to the bag does not decrease, so no vertex omitted from $C$ needs to be added.
If a new pair $u,v$ has distance at most $r$, then $a_X(u)\le r$ and~$u$ is represented in $C$.
Its distance to $v$ is determined by $\alpha_u$.
An already chosen landmark distinguishes the pair precisely when some $(p,\delta)\in\sigma_u$ satisfies
\[
 \delta\ne\min_{x\in X}\bigl(p(x)+\dist(x,v)\bigr).
\]
This tells us which pairs to add to $D$.
The information for $v$ itself is determined by $P$, because for a selected vertex $s$ with old profile $p$, the displayed minimum is exactly $\dist(s,v)-a_X(s)$.

We consider both choices for $v$.
If we select it, the size of the partial solution increases by one.
We add its profile to $P$ and add
$(p_{X\cup\{v\}}(v),\dist(v,u))$ to each $\sigma_u$. These distances are
coordinates of the vectors $\alpha_u$.
A pair is distinguished by $v$ if its two coordinates at $v$ differ, so we delete precisely these records from $D$.
In particular, all new pairs involving $v$ are deleted.
If the membership of $v$ is prescribed, we consider only the permitted choice.

Forgetting a vertex only changes the bag; the processed vertices and the chosen landmarks remain the same.
We restrict the recorded vectors to the new bag and normalize them again.
We discard a pair $(\alpha,\sigma)$ from $C$ if the minimum coordinate of its restricted $\alpha$ exceeds $r$.
If a landmark profile $p$ loses a coordinate and the minimum of the remaining coordinates is $c$, we replace each
$(p,\delta)$ by $(p-c,\delta-c)$, with the deleted coordinate omitted.
This accounts for the change in the distance of the landmark from the bag.

At a join node, the two processed parts intersect in their common bag $X$.
We combine two partial solutions only when they agree on which vertices of $X$ are selected.
Their union has size equal to the sum of their sizes minus the number of selected bag vertices, and its landmark profiles form the union of the two sets $P$.
A pair recorded by $(\alpha,\alpha')$ in one child is distinguished by a landmark from the other child precisely when one of that child's profiles $p$ satisfies
\[
 \min_{x\in X}(p(x)+\alpha(x))
 \ne \min_{x\in X}(p(x)+\alpha'(x)).
\]
We discard these pairs from the two sets $D$.

We must also consider pairs with one vertex in each child.
If such a pair has distance at most $r$, both vertices lie within distance $r$ of $X$, so their information occurs in the two sets~$C$.
For records $(\alpha_u,\sigma_u)$ and $(\alpha_v,\sigma_v)$ from the two children, separation by $X$ gives distance
\[
 h=\min_{x\in X}(\alpha_u(x)+\alpha_v(x)).
\]
If $0<h\le r$, the pair is distinguished by a selected vertex in the
first child precisely when some $(p,\delta)\in\sigma_u$ has
$\delta\ne\min_{x\in X}(p(x)+\alpha_v(x))$, and symmetrically for the
second child. If neither test succeeds, we add the jointly normalized
pair of vectors to $D$. Pairs involving bag vertices may be considered
more than once, which does not affect a set of records; the condition
$h>0$ excludes a vertex paired with itself.

For a vertex $u$ from one child, its distances to landmarks from the other child are again determined by their profiles at $X$.
We therefore add
\[
 \bigl(p,\min_{x\in X}(p(x)+\alpha_u(x))\bigr)
\]
to $\sigma_u$ for each landmark profile $p$ in the other child.
After making these additions to the records from both children, we take their union to obtain $C$.
The same distance calculations apply when one of the vertices belongs to $X$.

At a singleton leaf, the two possible choices can be checked directly, respecting any prescribed membership.
The calculations above show inductively that all table entries have the stated meaning.
As they depend only on the information stored in the tables, we may always keep a partial solution of smallest size for each entry.
At the root, all vertices have been processed, and a partial solution has the required property exactly when $D$ is empty.
The number of entries and the work needed to compute them at each node are bounded by a computable function of $w,\ell,r$.
Remembering which entries were combined allows us to recover a smallest set.
\end{proof}

\subsection{Landmarks in a ball}

% \todo[inline]{Is $S$ a resolving set of $G$ or $N_R(c)$?}

\begin{lemma}\label{lem:localized-resolving}
Suppose $G$ has a path decomposition of width at most $q$ whose bags
have diameter at most $q$. Given $G$, a vertex $c$, and a nonnegative
integer $R$, one can find a smallest resolving set $S$ \textcolor{black}{of the whole graph $G$} satisfying
\[
                         c\in S\subseteq N_R(c)
\]
in time $f(q,R)|V(G)|^{\mathcal O(1)}$, or determine that no such set exists.
% \textcolor{blue}{Only the locations of the landmarks are restricted to $N_R(c)$: the set $S$ must distinguish every pair of distinct vertices of~$G$, including pairs with one or both vertices outside the ball. All distances are measured in~$G$.}
\end{lemma}

\begin{proof}
Put $A=N_R(c)$ and $U=N_{R+q}(c)$.
We first show that a small set of vertices in $U$ separates the possible landmarks from the vertices outside $U$.
Consider a path decomposition as in the statement.
Since $A$ induces a connected graph, the bags meeting $A$ form an interval.
Each of these bags meets $A$ and has diameter at most $q$, so their union is contained in $U$.
Taking the first and last bags of the interval gives a set $Z\subseteq U$ of size at most $2q+2$ meeting every path from $A$ to $V(G)\setminus U$.
Notice that $Z$ may contain vertices of $A$.

We find a smallest such set $Z$ by a vertex-capacitated minimum cut,
giving capacity one to vertices of $U$ and infinite capacity to other
vertices. In the usual vertex-splitting construction, the source is
joined to the incoming copy of each vertex of $A$, and the outgoing
copies of vertices outside $U$ are joined to the sink. Thus vertices
of $A$ are themselves allowed in the cut. Let $C$ consist of $Z$ and
all components of $G-Z$ meeting $A\setminus Z$. Then
\[
 A\subseteq C\subseteq U,\qquad |Z|\leq2q+2,
\]
and every path from a vertex of $A$ to a vertex outside $C$ meets $Z$.
If $Z$ is empty, connectedness of $G$ implies $C=V(G)$.

We will express the conditions $c\in S\subseteq A$ and that $S$ resolves $G$ by a monadic second-order formula whose length, and whose number of additional unary colors, depend only on $q,R$.
Since $G$ has treewidth at most $q$, monadic second-order optimization will then find a smallest such set.
We mark $A$, $C$, and $c$ by unary colors.
Every vertex of $A$ has distance at most $M=2R+q$ from every vertex of $C$.
We can therefore express that $S$ distinguishes all pairs in $C$ using the formulas for distances $0,\ldots,M$.

It remains to describe the effect of the vertices outside $C$, whose distances from the landmarks may be arbitrarily large.
If there are such vertices, then $Z$ is nonempty.
For each of them write
\[
 \delta_v=\min_{z\in Z}\dist_G(v,z),\qquad
 \alpha_v(z)=\dist_G(v,z)-\delta_v.
\]
The diameter of $Z$ is at most $2R+2q$, so the coordinates of $\alpha_v$ belong to $\{0,\ldots,2R+2q\}$.
As $|Z|\leq2q+2$, the number of possible profiles depends only on $q,R$.
For $s\in A$ and a profile~$\alpha$, put
\[
 f_\alpha(s)=\min_{z\in Z}
                 \bigl(\dist_G(s,z)+\alpha(z)\bigr).
\]
Every path from $s$ to an outside vertex $v$ meets $Z$, and hence
\[
                    \dist_G(s,v)=\delta_v+f_{\alpha_v}(s).
\]
Observe that $0\leq f_\alpha(s)\leq M$.
Indeed, $\alpha$ has a zero coordinate, and the distance from $s$ to the corresponding vertex of $Z$ is at most $M$.
We use unary colors to record the value of $f_\alpha(s)$ for each $s\in A$ and each profile $\alpha$ that occurs.
These colors can be computed from the distance matrix of $G$, and their number depends only on $q,R$.

Consider first $x\in C$ and $v\notin C$.
If $\delta_v>M$, every vertex of $A$ distinguishes the pair, because its distance to $x$ is at most $M$.
Otherwise, the condition that a landmark distinguish~$x$ and $v$ depends on $v$ only through $\alpha_v$ and $\delta_v\in\{0,\ldots,M\}$.
For each combination that occurs, we require that for every $x\in C$ there be an $s\in S$ with
\[
                    \dist_G(s,x)\ne\delta_v+f_{\alpha_v}(s).
\]
The number of these requirements depends only on $q,R$, and each is expressible using the colors above and bounded-distance formulas.

Now consider two distinct vertices $v,w$ outside $C$.
If $|\delta_v-\delta_w|>M$, every vertex of~$A$ distinguishes them.
Otherwise, their two profiles and the difference $\Delta=\delta_w-\delta_v\in\{-M,\ldots,M\}$ determine which vertices of $A$ distinguish them: we need an $s\in S$ for which
\[
                  f_{\alpha_v}(s)-f_{\alpha_w}(s)\ne\Delta.
\]
For each combination realized by a pair of distinct outside vertices, we mark the vertices of~$A$ satisfying this inequality and require $S$ to contain one of them.
If no vertex of $A$ satisfies the inequality, no resolving set contained in $A$ exists.
Again, the number of color classes is bounded in terms of $q,R$.

We have now expressed that every pair of vertices of $G$ is distinguished by $S$.
The formula has length bounded in terms of $q,R$, as does the number of colors, and all colors and realized combinations can be computed in polynomial time.
Unary colors do not affect treewidth.
We can therefore use Bodlaender's algorithm~\cite{Bodlaender1996} to find a tree decomposition and apply monadic second-order optimization~\cite{ArnborgLagergrenSeese1991} to find a smallest satisfying set in the claimed running time.
\end{proof}

\subsection{Putting the algorithm together}

\begin{proof}[Proof of \cref{thm:path-fpt}]
A one-vertex graph has the empty resolving set, so suppose $|V(G)|\geq2$ and let $q=\bcpw(G)\geq1$.
Since $\ctw(G)\leq q$, \cref{lem:short-tree-decomposition} gives a tree decomposition whose width and bag diameter are both $\mathcal O(q^2)$.
By \cref{lem:find-dominating-geodesic}, we can also find a $q$-dominating shortest path.

For each pair $s,t$ with $\dist(s,t)>6q$, we apply \cref{lem:local-resolving-dp} with $r=6q$, requiring both $s$ and $t$ to be selected.
The resulting set resolves all nearby pairs by construction, and all other pairs by \cref{lem:distant-landmarks}.
For every vertex $c$, we also apply \cref{lem:localized-resolving} with $R=6q$.
Among all sets found in this way, we return one of smallest size.

Every set considered for the output is a resolving set.
Conversely, let $S$ be a minimum resolving set.
If it contains two vertices farther than $6q$ apart, the computation for this pair returns a set of size at most $|S|$.
If it does not, choose any $c\in S$; the set is nonempty because the graph has at least two vertices.
Then $S\subseteq N_{6q}(c)$, and the computation for $c$ returns a set of size at most $|S|$.
Hence the algorithm finds a minimum resolving set.
There are at most $|V(G)|^2$ computations, each taking $f(q)|V(G)|^{\mathcal O(1)}$ time.

Finally, the algorithm need not know $q$ in advance.
The bag-connected tree decomposition is computed once, independently of $q$, and reused throughout, including for the monadic second-order optimization.
We try $q=1,2,\ldots$ until we have found the structures used in the proof.
We check that the shortest path found is $q$-dominating and that the minimum cut for every ball of radius $6q$ has size at most $2q+2$.
If one of these conditions fails, we again increase $q$.
Once they all hold, the correctness proof applies: it only uses these structures, not a path decomposition witnessing $\bcpw(G)\leq q$.
All conditions hold by $q=\bcpw(G)$, and the computed decomposition has width and bag diameter $\mathcal O(\bcpw(G)^2)$, so the total running time has the required form.
\end{proof}

In particular, bounded width with connected bags gives polynomial-time solvability for path decompositions, whereas \cref{thm:main} rules it out for tree decompositions unless $\mathrm P=\mathrm{NP}$.
The connected-prefix requirement is different: \cref{lem:pathwidth-four} already gives hardness at width four under that convention.

\section{Acknowledgements}
We thank Florent Foucaud, Oscar Defrain, Prafullakumar Tale, Mathieu Mari for many interesting discussions on the topic. We acknowledge the organizers and participants of the June 2026 GRALMECO-GRAPHMETRIX Workshop in Aydat near Clermont-Ferrand, France, where many interesting discussions were carried out.

\begingroup
\small
\bibliographystyle{plainnat}
\bibliography{Metric_Dimension_Connected_Treewidth_Hardness}
\endgroup
\end{document}